\documentclass[a4paper, 11pt]{article}
\usepackage{amsmath}
\usepackage{subfigure}
\usepackage{graphicx}
\usepackage{bm}
\begin{document}
\title{Failures and avalanches in complex networks}
\author{Jan {\O}ystein Haavig Bakke$^1$ \and Alex Hansen$^1$ \and J\'anos Kert\'esz$^2$\\
  {\small 1: Department of Physics, Norwegian University of Science and Technology}\\
  {\small 2: Institute of Physics, Budapest University of Technology}
}


\maketitle

\begin{abstract}
We study the size distribution of power blackouts for the Norwegian and
North American power grids.  We find that for both systems the
size distribution follows power laws with exponents $-1.65 \pm 0.05$ and 
$-2.0 \pm 0.1$ respectively. We then present a model with global redistribution
of the load when a link in the system fails which reproduces the power law
from the Norwegian power grid if the simulation are carried out on the Norwegian high-voltage 
power grid. The model is also applied to regular and  irregular networks and give power laws with 
exponents $-2.0\pm0.05$ for the regular networks and $-1.5\pm0.05$ for the irregular 
networks. A presented mean field theory is in good agreement with these numerical results.
\end{abstract}

Large transportation networks like the road system, pipelines 
and the electrical power grid are sensitive to local failures. 
When failures occur the transport on these networks must be 
redistributed on the still intact part of the network, occasionally exceeding
the local capacity and causing further failures\cite{lgkk05,scl00}. The 
resulting avalanches may finally end in major breakdowns: megajams in 
vehicular traffic or blackouts in the electrical power system. 

Such systems get increasingly complex with time as they and the 
transport on them grow. In addition, the liberalisation of the 
electrical power distribution market adds more complexity to the 
power grid system, since the market mechanisms give feedback to the 
production and transport of power over large regions. As the 
complexity of these systems increases  the ability to predict the 
behaviour of large unwanted events becomes more and more difficult.

Studies of vulnerability and avalanche statistics in complex networks have been done
using different models. In the Motter and Lai model \cite{ml02} the load on the network 
is described by betweenness centrality. Avalanches generated by this model on scale free 
networks were found to follow power law distribution \cite{lgkk05}. Alberts et al. studied 
a similar model on the North American power grid.\cite{aan04}

Transport properties characterised by conductance for different networks have also
been studied on a general basis.\cite{lbhs05} And the blackouts in power grids have 
also been studied in the context of self-organised criticality\cite{scl00}. 

The aim of this Letter is to present a model for such systems 
using the network of the electric power grids and to compare 
the results with observed data. The electric power distribution 
system, or power grid, is a system of 
generators, transformers, power lines and distribution substations.  
Earlier studies of the North American power grid have found the degree distribution
$P(k) \propto e^{-0.5 k}$ with mean degree of $2.67$ 
\cite{aan04, ws98}. We find that the Norwegian power grid also has an 
exponential degree distribution, but with a mean degree of $2.28$. 
The number of nodes in he North American power grid is of the order of 15000 nodes 
(generators, substations,...) whereas the Norwegian one is of the order 
of 1000 nodes. The difference in mean degree might reflect the fact that 
Norway is sparsely populated, and have large  regions that are only supplied 
by one or two major grid lines, leading to a smaller value of the mean degree. 
Power grids are thought to have small  world properties, i.e., a large degree of 
clustering and small characteristic  length and  where the characteristic path 
length $l$ scale logarithmically with the number of nodes \cite{ws98,n03}. 

The failure distribution has already been studied for the North 
American power grid \cite{cndp00}.  The 
statistics of the failures will depend on which quantity is 
used to characterise the size of a failure, either the energy 
unserved or the power lost.  However, both quantities lead 
to power laws.  We will in the following compare the model 
we present in this Letter to the power loss distribution 
both for the Norwegian and the North American power 
grids \cite{nxx,mxx}. We believe that this is the better of 
the two quantities to use, as the energy unserved will 
depend on human factors like how long it took to repair a 
given transmission segment that has broken down.   

We present the possibly simplest model which is capable of
describing the above mentioned avalanche effects. Imagine a 
network consisting of electrical conductors, all with the same 
conductance.  One node is picked at random and current will 
be injected here.  Another node, different from the first, will 
act as a current drain. A potential difference is set up 
between these two nodes and the Kirchhoff equations are solved to 
find the current in each link\cite{bh88}.  Once the currents $i$ have 
been found, we assign a breakdown threshold $t$ to each 
link such that $t=(1+\alpha)i$, where $\alpha$ is a 
positive number.  The idea here is that each transmission 
line has been constructed with a fixed tolerance $\alpha$. 
We found that having  more current source/sink pair gave
similar results as those found for smaller network, thus effectively reducing the
system size.

We now choose a link and remove it in the same way as is done in the 
random fuse model \cite{arh85,hr90,h05}. 
The currents are then recalculated. This gives global rearrangement in 
contrast to the models of  self-organised criticality \cite{pb99} where  load 
rearrangement is local. Some  links will at this point carry a current 
{\it larger\/} than their threshold.  These links are then removed from 
the network and the currents are recalculated once more.  Again, the bonds 
where the current exceeds the threshold are removed  and this procedure is 
repeated until all currents are below the thresholds.  This removal process 
models an avalanche of failures.  

Let the conductance between the source and the sink before the initial random 
removal of a link be denoted by $G_i$ and similarly, let $G_f$ be the 
conductance after the avalanche is finished. The difference $\Delta G = 
G_i-G_f$ is a measure of the magnitude of the avalanche generated by that 
removal. If the voltage $V$ between the source and sink 
nodes is kept fixed during the blackout event, the change in conductance 
$\Delta G$ will be proportional to the power loss: 
$\Delta P=P_i-P_f=V^2\Delta G$. For normalisation 
purposes, we will in the following define $\Delta G= (G_i-G_f)/G_i$.  

The results we present here are based on generating ensembles of networks and
for each network systematically choosing every link as initiator of a
blackout event.  We study two classes of networks: Random networks 
with exponential degree distribution, $P(k) \propto e^{-0.5k}$  which is the same as for
the Norwegian and North American networks, and small world networks \cite{nw99} with mean degree 
$2.67$. Both these network classes have a mean shortest distance $l$ that scale 
logarithmically with the system size.

We also study square and triangular lattices with bi-periodic boundary conditions,
 where $l$ grows as the square root of the number of nodes. In addition to 
these networks and lattices we implement the model on networks with the 
topology from the Norwegian and North American power grid. 

We did simulations for network sizes up to 5041 nodes for the four different 
artificial network types described above, using the conjugate gradient algorithm 
to solve  the Kirchhoff equations on the network using an error-criterion of 
$10^{-16}$ \cite{bh88}. We find that the random and the small world networks 
give power law probability distributions for the conductance loss 
$p(\Delta G)\sim \Delta G^{-\gamma}$, as does simulations on the network
based on the Norwegian power grid.  The regular lattices also give rise to
power laws, but with a different value for  $\gamma$.
It is worth noting that in a study of
a somewhat related model by Roux et al.\ \cite{rhh91}, power laws were 
observed.  In that model, a square lattice between two parallel bus bars
are gradually depleted by the randomly removing link by link and recording
in a histogram the conductance changes between 
each removal. The scaling of the conductance changes 
could be traced back to the increasing connectivity length during the process. 
In this depletion process the thresholds are uncorrelated with the initial
currents, while in our case the threshold and initial current distributions
are identical.

\begin{figure}
  \includegraphics[scale = 0.4]{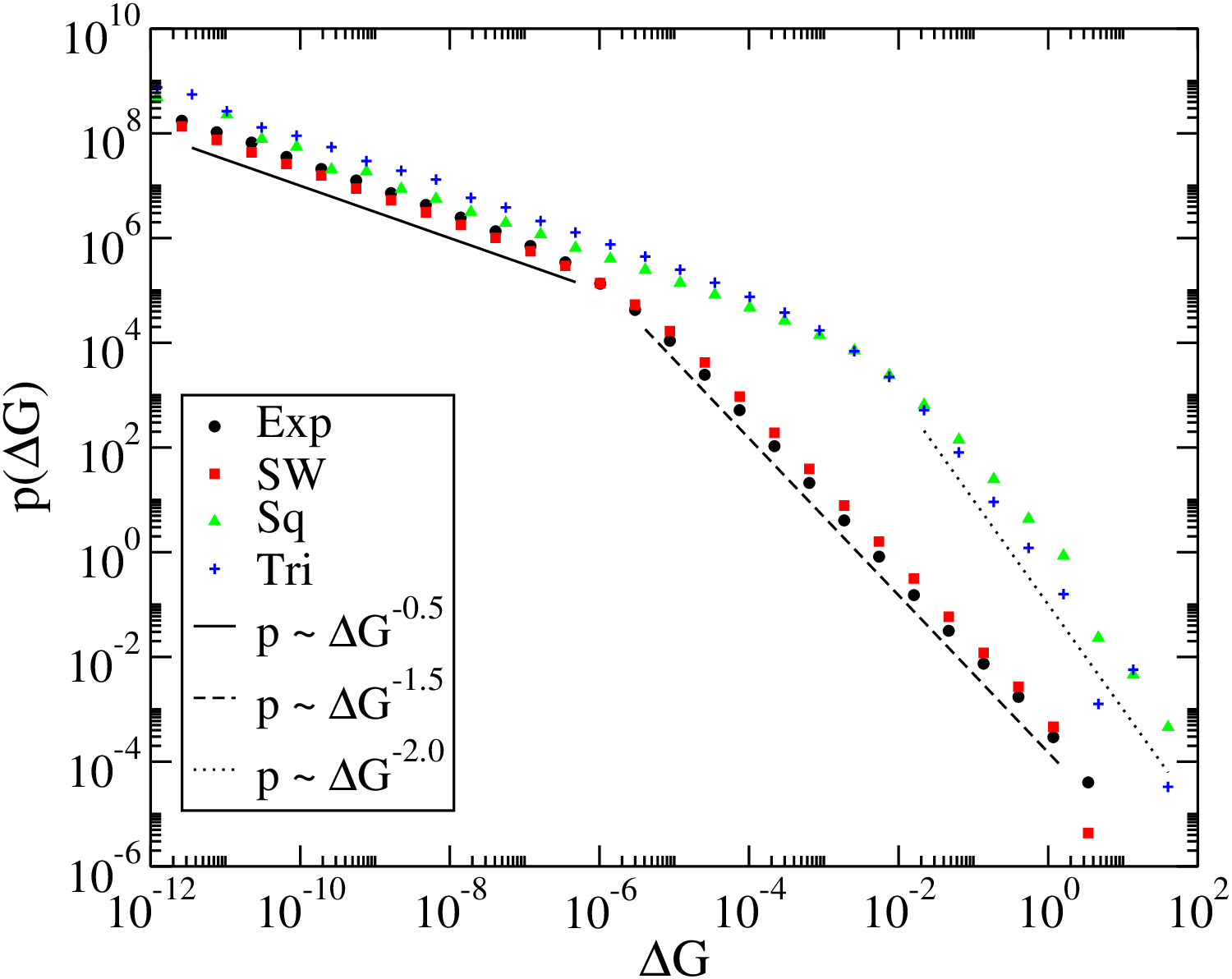}
  \caption{\label{fig1}Probability density function of 
    conductance losses for different network and lattice types. The data for 
    the regular lattices have been moved two orders of magnitude towards the right 
    to separate them from the irregular networks. (Exp = exponential network, SW = 
    small world network, Sq = square lattice, Tri = triangular lattice). Some data
    points for the irregular network have $\Delta G > 1$ since the data is put at
    the midpoints of the logarithmic bins.}
\end{figure}

\begin{figure}
  \includegraphics[scale = 0.4]{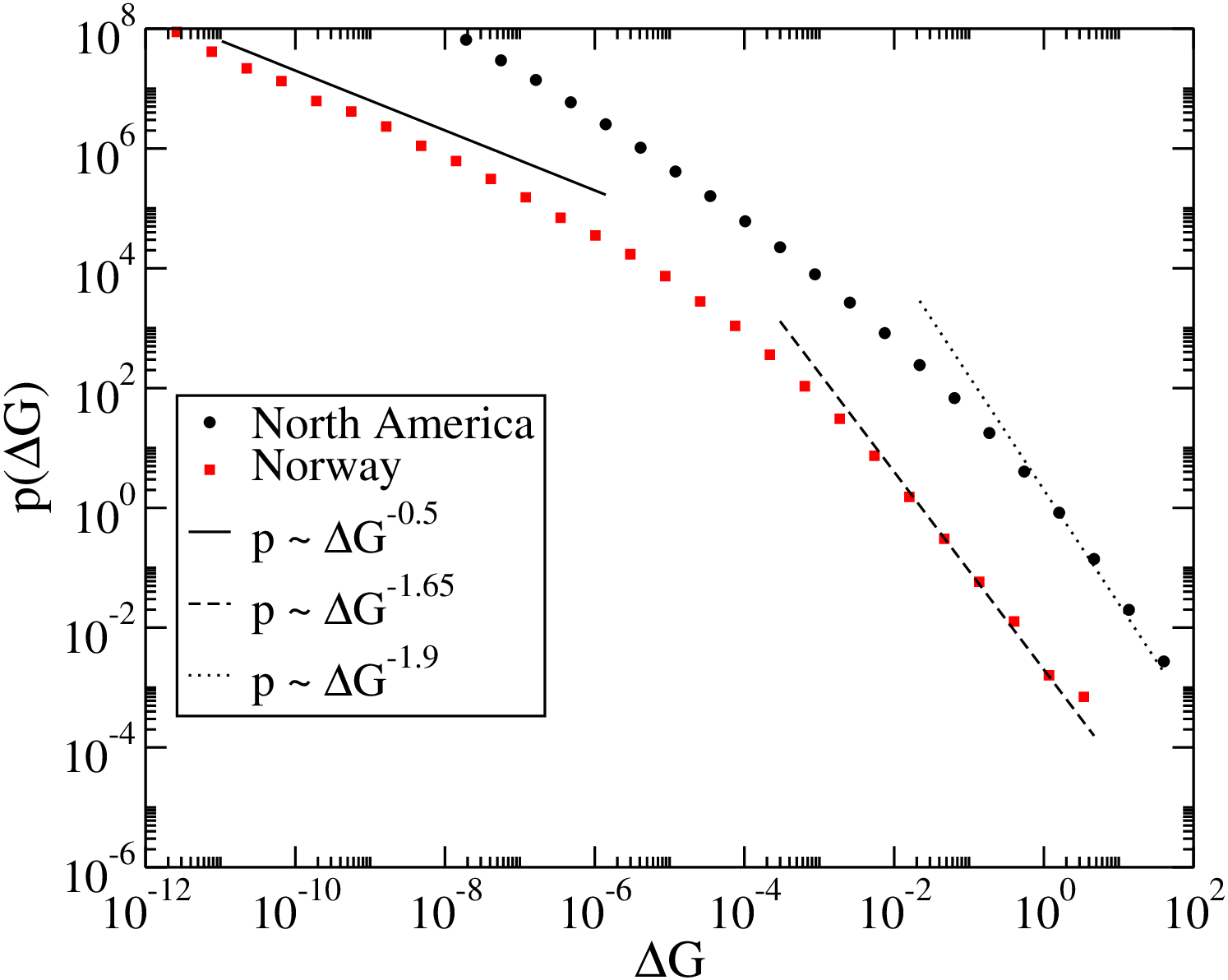}
  \caption{\label{fig5} Our model implemented on the Norwegian (1220 nodes) and North
    American Power (4941 nodes) grids.  The straight lines correspond to the power laws
    observed in the data of fig.\ \ref{fig4b}.}
\end{figure}

We show in fig.\ \ref{fig1} the probability density of 
conductance changes $\Delta G$
for the random exponential, the small world networks and 
for the two regular networks. We find that the conductance 
changes follow power laws in two different regimes.  There is a small-event
regime characterised by $p(\Delta G) \sim \Delta G^{-0.5}$.  The exponent
$-0.5$ seems to be independent of network or lattice type. 
The large-event regime is characterised by two different power laws.  The 
irregular networks follow a power law characterised by an exponent 
$-1.5\pm0.05$, whereas the regular lattices follow a power law characterised
by an exponent $-2.0\pm0.07$. 

The simulations were performed with 
$\alpha = 3.0$. For smaller values of $\alpha$ a large number of the breakdowns
broke the system completely with $\Delta G = 1.0$, thus destroying the power law 
tail for large $\Delta G$s. Larger values of $\alpha$ did not change the tail of
the distribution. For the real power blackout data the largest events  that were 
recorded removed $5$\% of the totalt capacity of the system. This fact supports the 
use of an $\alpha$ that does not break the system completely i. e. $\Delta G < 1$.

For both the random and the  small-world networks, we find that the initial 
current distribution follows a power law, hence giving a power law distribution
of the thresholds.

We now present a mean field estimate of the current distribution in an infinite
network.  Assume that $n(r)$ is the average number of links at a distance
$r$ from an arbitrarily chosen origin. $r$ is the graph theoretical distance, which is
of the same order of magnitude as the Euclidean one for a lattice.  
If a current is injected into the network at the origin, the typical current $i(r)$ 
in a link at a distance $r$ from the origin will be inversely proportional to the 
average number of links at that distance, $i(r)\propto 1/n(r)$.  Since $n(r)$ is 
a monotonically increasing function of $r$, $i(r)$ is a monotonically decreasing 
function of $r$: The smaller the $r$, the larger the average current $i(r)$. 
The number of links carrying a current higher than a given value
$i$, i.e., the cumulative current distribution $P(i)$, is then
simply given by
\begin{equation}
  P(i) \sim \int_0^{r(i)} n(r') dr',
  \label{eq.area}
\end{equation}
where we have defined $r(i)$ as the solution with respect to $r$ of 
the equation $i = 1/n(r)$. 

\begin{figure}
  \includegraphics[scale = 0.4]{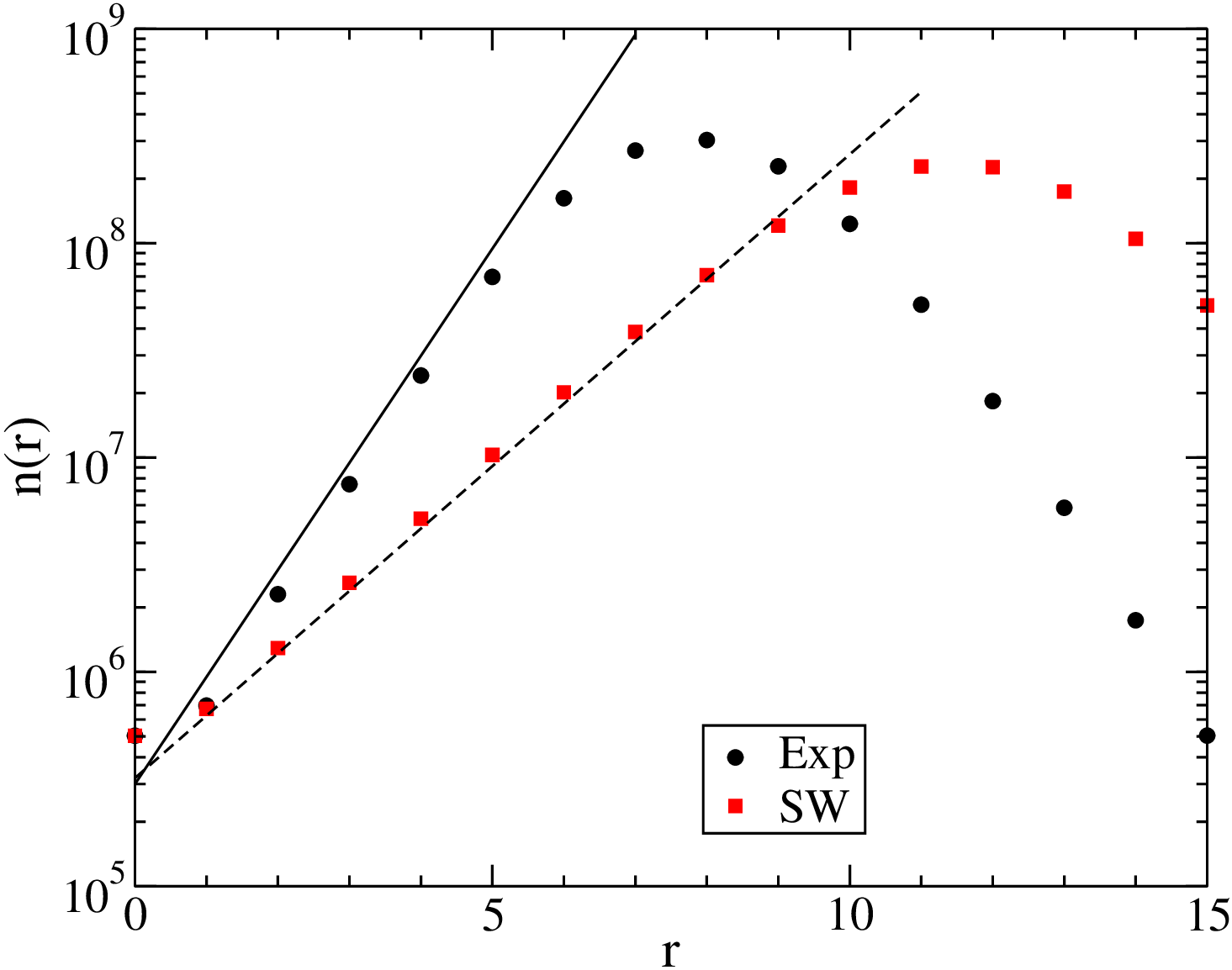}
  \caption{\label{fig2} The average number of neighbours $n(r)$ at a distance $r$ for the
    exponential and the small world networks. $n(r)$ falls of for large $r$ due to 
    the finite size of the networks.}
\end{figure}

\begin{figure}
  \includegraphics[scale = 0.4]{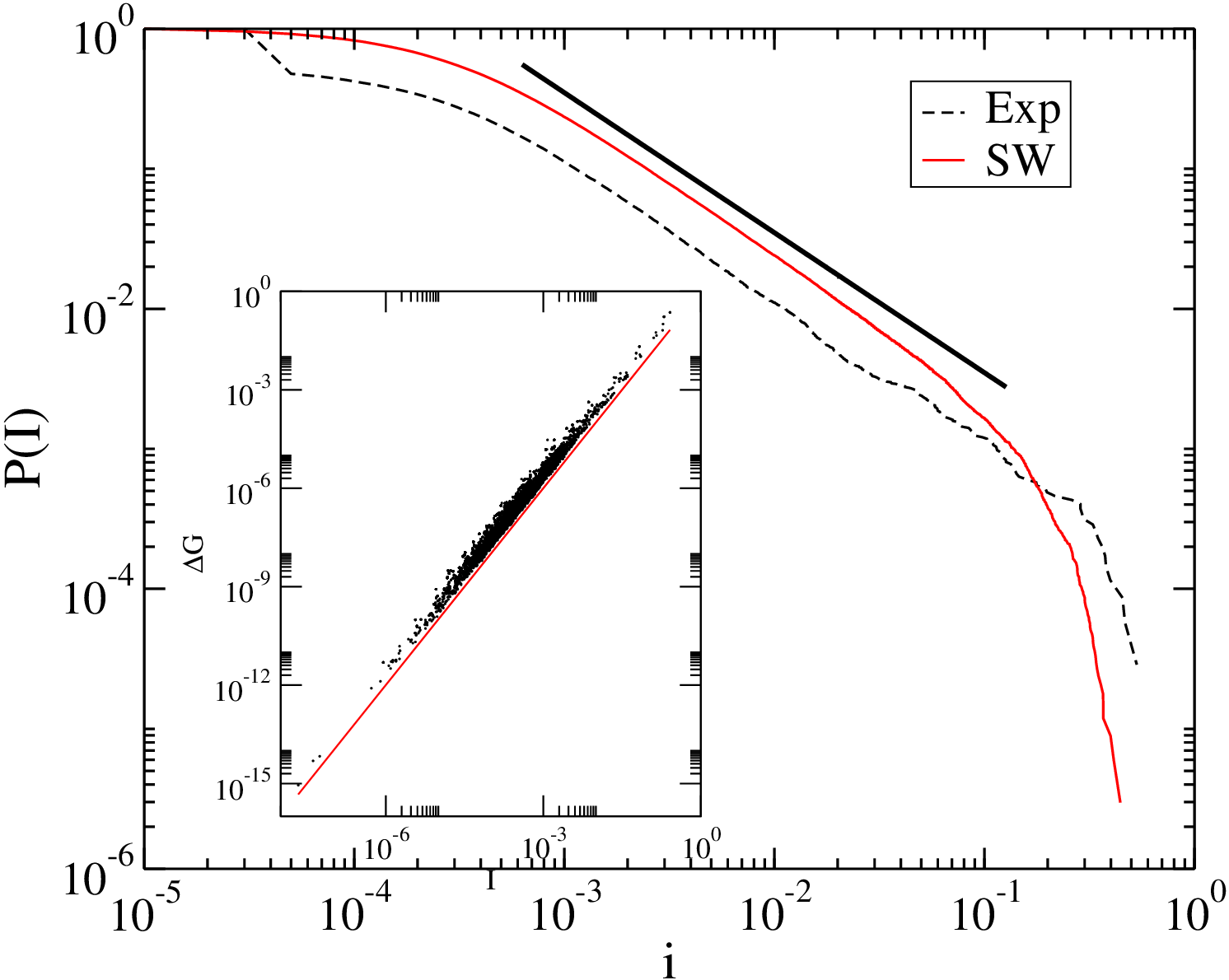}
  \caption{\label{fig3} Cumulative current distribution for 
    random exponential and small scale networks each containing 5041 nodes. 
    The distribution function is  $P(i) \sim i^{-1}$, as excepted from eq.\ (\ref{eq:pnr}).
    The inset shows $\Delta G$ vs. $i$. for $\alpha = 3.0$. The solid line in the 
    inset is $i^2$.}
\end{figure}

For the random exponential networks and the small world networks we have
that $n(r) \sim  e^{\beta r}$ for small $r$.  This is shown numerically in
fig.\ \ref{fig2}.  Hence, combining this behaviour with
eq.\ (\ref{eq.area}), we find
\begin{equation}
\label{eq:pnr}
P(i) \sim \int_0^{-(1/k)\ln(i)} e^{\beta r'}dr'\sim \frac{1}{i},
\end{equation}
for large $i$.  This is the behaviour we see in fig.\ \ref{fig3}.

We note here, as support for the mean field argument we have just presented, 
that for square and triangular networks, where 
$n(r)\sim 1/r$, the mean field argument gives 
$P(i)\sim i^{-2}$, which we also verified numerically.

Assume that the resistance $\Delta R$ when a link is broken  is much less than the total resistance $R$.
With constant voltage difference between the sink and source $\Delta R$ is proportional to $\Delta G$
to the first order of $\Delta G$
\begin{equation}
  \Delta R \sim \frac{\Delta G}{G^2},
  \label{eq:dRdG}
\end{equation}
where $G$ is the conductance for the whole system.

Assuming an intermediate length $l$, larger than the lattice constant, but smaller than
the system size, Roux et al.\ \cite{rhh91} argued using Tellegen's theorem 
from network theory in electrical engineering \cite{psd70} that 
\begin{equation}
  \Delta R = \Delta R_l \left( \frac{i_l}{I} \right)^2
  \label{eq:tell}
\end{equation}
where $\Delta R_l$ is the resistance change at scale $l$, $i_l$ is the current of a
region of scale $l$ and $I$ is the total current.
Combining eq.  (\ref{eq:dRdG}) and eq.\ (\ref{eq:tell}) one gets
\begin{equation}
\Delta G_j \sim i_j^2
\label{eq:roux}
\end{equation}
which describes the relation between the conductance loss to the system when bond $j$
is removed and $i_j$ is the current through the bond. We show the correlation between
conductance change $\Delta G_j$ and the current $i_j$ during the breakdown
process for our model in the inset of fig.\ \ref{fig3}.  

With the cumulative current probability shown in fig. \ref{fig3} and using 
eqs.\ (\ref{eq:pnr})  and (\ref{eq:roux}) we would expect a distribution function 
$p(\Delta G) \sim \Delta G^{-1.5}$ for the exponential and small world networks, 
which is indeed what was observed in fig. \ref{fig1}. We also find that the above argument 
also predicts the correct distribution function for the regular networks in fig.\ \ref{fig1}, 
$p(\Delta G) \sim \Delta G^{-2.0}$.

We note that this argument is based on each breakdown event only involving a single
bond.  If there is a typical or dominating current carried by the bonds in 
the avalanches involving more than one bond --- and a corresponding 
conductance change, the argument we have presented carries over to this situation. 

The power law for small events, $p(\Delta G)\sim\Delta G^{-0.5}$ in 
fig.\ \ref{fig1} corresponds to a uniform distribution function for
the corresponding currents, when using eq.\ (\ref{eq:roux}).  

Finally, we compare the results from our simulations with blackout data 
from the 
Norwegian main power grid and data from the largest blackouts in the North 
American power grid \cite{mxx,nxx}. These data are different due 
to the fact that the North American data set only look at large events 
$\Delta P > 10$ MW, while the Norwegian data set is for $\Delta P > 0.1$ MW. 

\begin{figure}
\subfigure{\label{fig4a}
  \includegraphics[scale = 0.4]{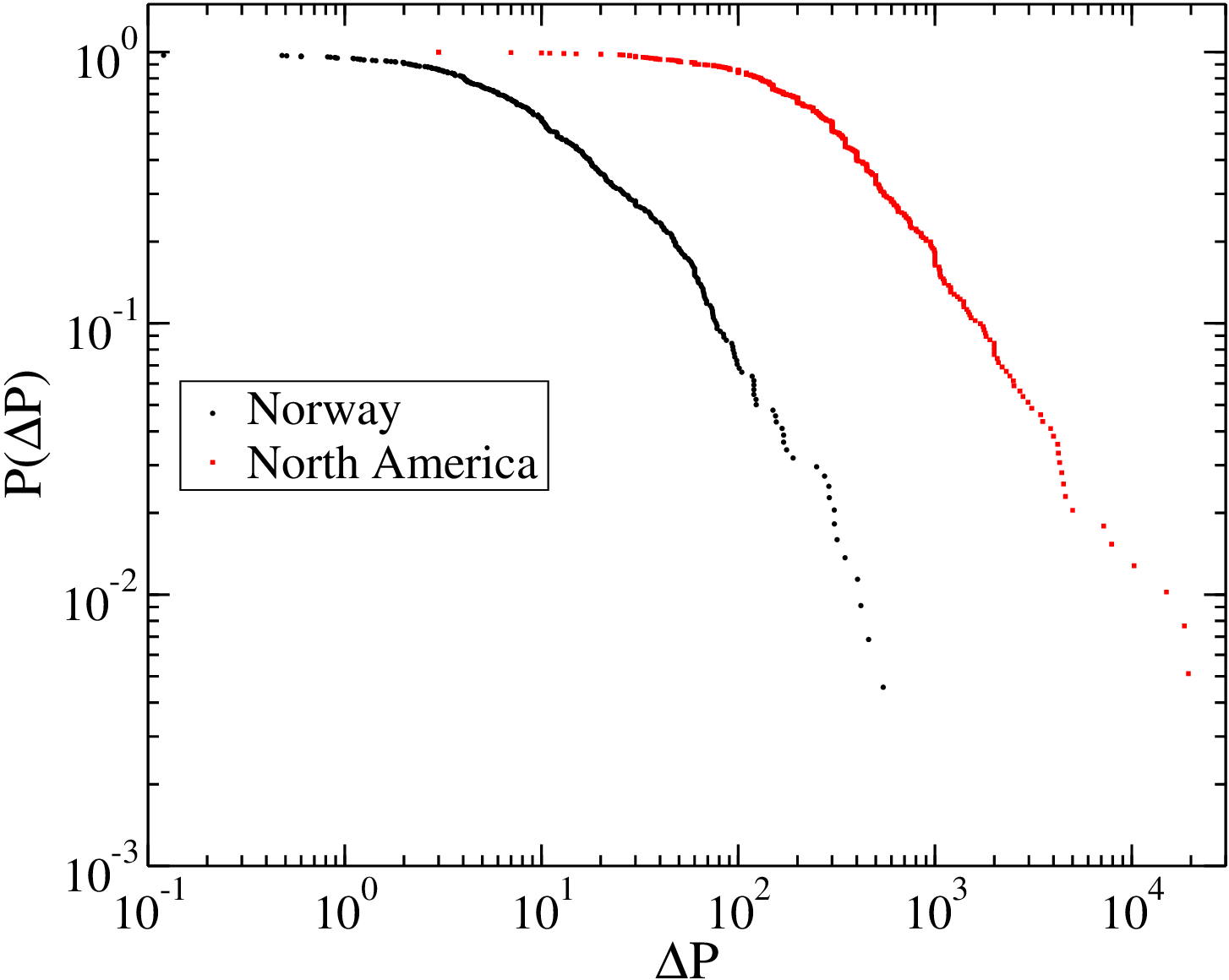}}
\subfigure{\label{fig4b}
  \includegraphics[scale = 0.4]{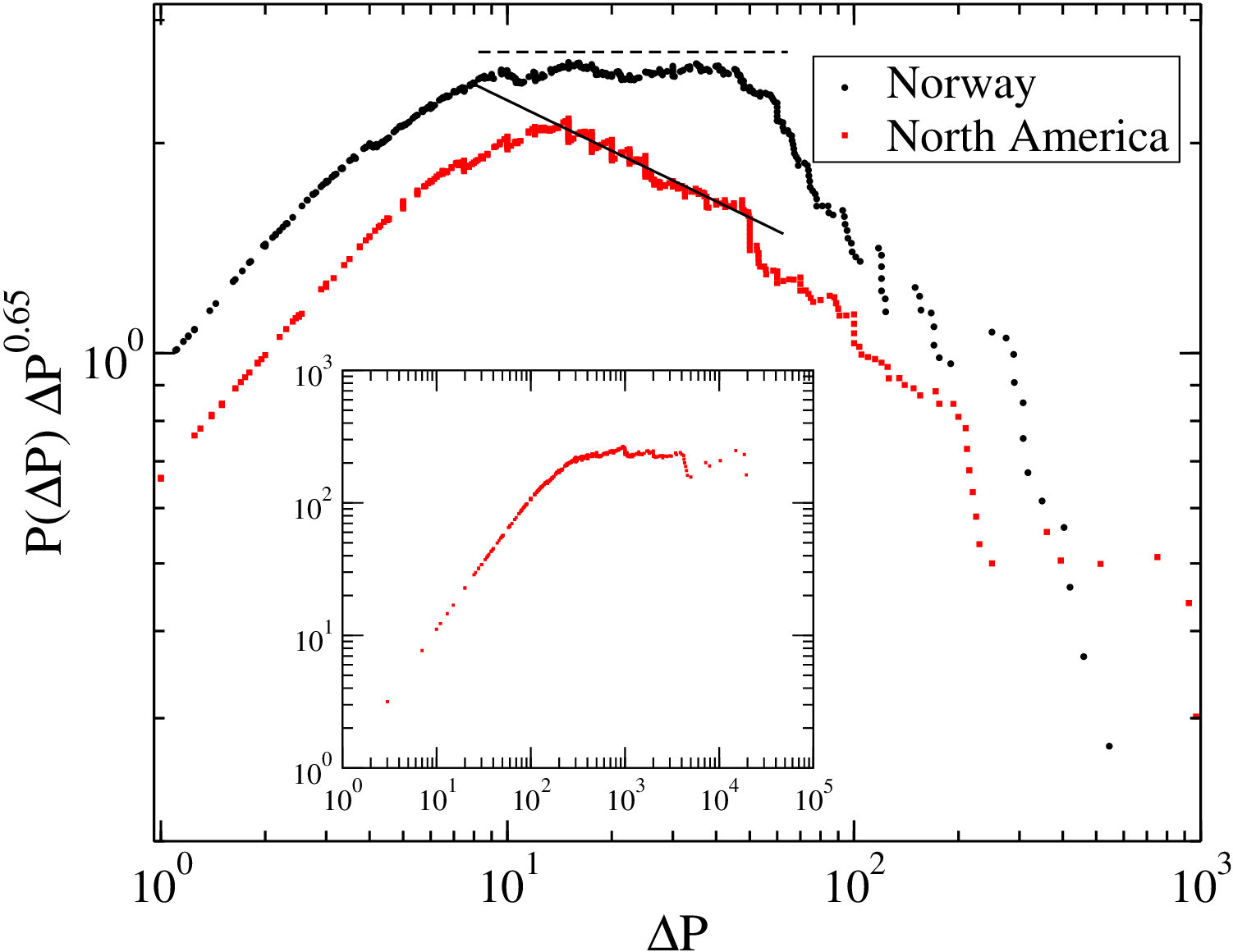}}
\caption{\label{fig4} a) Cumulative distribution $P(\Delta P)$ for blackout
events $\Delta P$ in the Norwegian (373 events) and North American (390 events) power grids.
b)  $P(\Delta P)\Delta P^{0.65}$ for blackout
events $\Delta P$ in the Norwegian and North American power grids.
The straight line to the left following the North American
data has an exponent of $-0.24$.}
\end{figure}

The power blackout data from the main Norwegian central power grid was collected for 
the period 1995-2005. The North American data span the period 1984-2002. In 
fig.\ \ref{fig4a}, we show the cumulative probability $P(\Delta P)$ giving the
probability to find an event larger than or equal to $\Delta P$.  This
function is extracted from the data by ordering them in a ascending sequence
and then plotting event $k$ in the sequence along the abscissa together with
$k/(N+1)$ along the ordinate, where $N$ is the total number of events
\cite{d81}.

In fig.\ \ref{fig4b}, we have shifted the data North American data from
 fig.\ \ref{fig4a} to simplify the comparison of the data. The cumulative 
probability has furthermore been multiplied by $\Delta P^{0.65}$. The 
ensuing flat plateau in the Norwegian  1995--2005 data suggests that the 
probability density follows a power law of the form 
$p(\Delta P) \sim \Delta P^{-1.65}$. For large values of $\Delta P$ it falls off 
faster. The North American data do not show such a plateau.  However, they are 
consistent with a law regime corresponding to  $p(\Delta P) \sim \Delta P^{-1.9}$. 
The inset shows  $P(\Delta P)\Delta P^{1.05}$ which indicates a power law of the 
form  $p(\Delta P) \sim \Delta P^{-2.05}$ for the North American data supporting
a exponent $-2$ for this blackout distribution without a cut-off for large $\Delta P$s.

We show in fig.\ \ref{fig5} an implementation of our model on the Norwegian
and North American power grids\cite{grids}.  We have fitted power laws to the distribution
with exponents $-1.65$ and $-1.9$.  These are the power laws that were observed
in the blackout histograms shown in fig.\ \ref{fig4b}.  We see that the model
produces data that are consistent with the observations for the Norwegian 
power grid, while the data for the North American power grid is inconclusive.  
It is furthermore interesting to observe that the exponents occurring in the 
data for moderately sized blackouts lie in between the results of the model 
implemented on  the irregular networks (exponent -1.5) and the regular 
lattices (exponent -2). The model does not reproduce the large scale blackout 
distributions for the Norwegian power grid which fall of faster than
$\Delta P^{-1.65}$. We also see a small scale regime as for the artificial 
networks in fig.\ \ref{fig1}. Hence, the simple model we have introduced is 
capable to reproduce some aspects of the observed blackout distribution 
quantitatively with reasonable precision.  

The difference for large $\Delta P$s in the Norwegian and the North American 
datasets could be accounted for by the fact that the North American data also 
includes large events like snowstorms, hurricanes, while the Norwegian data do 
not include these events. The difference in nature of widespread events like a 
hurricane compared with a power line fault can be the cause of the cut off in the 
Norwegian dataset. This could also be the reason why the exponent for the Norwegian 
power grid is close to the theoretical value for irregular networks since this argument is
based on breaking single links. 

The reason why there is a difference in the conductance loss distributions for the simulation with the 
Norwegian and the North American power grid is not clear. There are however some differences 
between these two networks when we look at more than just the degree distribution. We observed 
two differences in $n(r)$ for these to power grids.  First $n(r)$ for the Norwegian network 
is closer to an exponential than the North American network for small $r$ values, 
and there is a more pronounced peak in $n(r)$ for the Norwegian  network, while the 
North American network have relatively wide plateu. This could explain the 
difference in $p(\Delta G)$ found in the simulations done on these networks.

We thank Statnett and R.\ K.\ Mork for providing us with the Norwegian breakdown data
| and NVE and A.\ E.\ Gr\o nstvedt for data on the Norwegian power grid. J.\O .H.B.\ 
thanks Ingve Simonsen for discussions on creating probability distributions. This 
work was  partially supported by OTKA K60456  and the Norwegian Research Council.
 A.H.\ thanks the Collegium Budapest for hospitality 
during the period when this work was initiated. 



\begin{thebibliography}{99}
\bibitem{mxx}Data for the Norwegian blackouts was provided by Statnett.
\end{thebibliography}
\end{document}